\documentclass[aps,prc,twocolumn, superscriptaddress,showkeys,nofootinbib]{revtex4-1}  

\usepackage[utf8]{inputenc}
\usepackage{graphicx}
\usepackage{subcaption}
\usepackage{graphicx,color}
\usepackage{amsmath,amssymb,amsfonts}
\usepackage{hyperref}

\usepackage{xspace}	

\begin{document}
\title{
Charged-Current Elastic Scattering at the Electron-Ion Collider
}
\medskip
\author{Henry~T.~Klest} 
\email{hklest@anl.gov}
\affiliation{Physics Division, Argonne National Laboratory, \\
Lemont, IL 60439, USA}

\begin{abstract}
We discuss the measurement of the charged-current elastic scattering process $e^-p\rightarrow\nu_e n$ at the Electron-Ion Collider (EIC). This process provides sensitivity to the poorly constrained axial radius of the nucleon. Collisions of electrons with polarized protons enable measuring the $e^{-\!}\,\vec{p} \to \nu_e\,n$ target-spin asymmetry for the first time. We conclude that a measurement of charged-current elastic scattering at the EIC will, perhaps unsurprisingly, prove very challenging. However, with dedicated instrumentation at a second EIC detector, the measurement may be possible.
\end{abstract}
\maketitle

\section{Introduction} \label{sec:intro}

Elastic scattering on nucleons and nuclei has been one of the primary tools of nuclear and hadronic physics for over a century. The momentum-transfer dependence of elastic scattering processes provides a path to characterize the structure of the target, and different elastic processes can provide complementary information about the distributions of quantum numbers, such as electric charge, mass, and even spin~\cite{Hofstadter1956,PolyakovSchweitzer2018,Miller:2019DefiningRp,Chen:2024oxx,KaiserWeise:2024Sizes,Goharipour:2025yxm}. 

\begin{figure}[h]
    \centering
    \qquad 
    \includegraphics[width=0.5\linewidth]{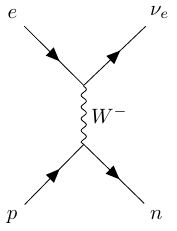}
    \caption{Diagram of the CCE process. In the elastic process, unlike charged-current deep inelastic scattering, the $W^-$ does not resolve the partonic structure of the nucleon and the process is sensitive to the axial form factor.}
    \label{fig:CCE}
\end{figure}

The elastic process we will focus on in this paper is the charged-current elastic (CCE) scattering process, $e^-p\rightarrow\nu_e n$ (Fig.~\ref{fig:CCE}), whereby the structure of the proton is observed through the lens of a $W^-$. The fact that in this process the $W^-$ nominally couples only to left-handed up-type quarks in the proton means that it serves as a unique filter for studying flavor- and spin-dependent contributions to the nucleon structure. The form factor associated with this process is the axial form factor\footnote{Also called the nucleon axial-vector form factor; some literature also uses ``nucleon transition axial form factor", since technically it describes a transition from $n\rightarrow p$ or vice versa.}, $F_A(t=-Q^2)$\footnote{In elastic scattering, the momentum transfer to the proton and the lepton are equivalent. Therefore, $|t|
$ and $Q^2$ can be used interchangeably, and we use only $|t|$ for convenience in the following text.}~\cite{LlewellynSmith1972}. This form factor encodes the momentum-transfer dependence of the nucleon matrix element of the axial current. The so-called nucleon ``axial charge radius"\footnote{Throughout, we use the term ``radius'' in the standard form-factor sense, where the radius is defined via the slope of the form factor at $|t|=0$. In a relativistic composite system this need not coincide with a unique, frame-independent three-dimensional static density~\cite{Miller:2019DefiningRp}.
}, $r_A$, is defined via 
\begin{equation}
\langle r_A^2 \rangle \equiv-\frac{6}{F_A(0)}\frac{\mathrm{d}F_A}{\mathrm{d}|t|}\big|_{|t|=0},
\end{equation} 
where $F_A(0)$ refers to the value of the form factor at zero momentum transfer~\cite{MINERvA:2023avz}. $r_A$ encodes the behavior of axial currents globally in the nucleon, and measurement of $r_A$ is an attractive goal for future experiments.  The authors of Refs.~\cite{Chen:2024oxx,Chen:2024ksq} highlight $r_A$ as the primary component of a proton ``spin radius", defined as 
\begin{equation}
\langle r_{\mathrm{spin}}^2 \rangle = r_A^2+\frac{1}{4M^2}(1+\frac{2F_p(0)}{F_A(0)}),
\end{equation}
where $F_p(0)$ is the induced pseudoscalar form factor at zero momentum transfer.

The value of $F_A(0)$ is known to high precision from $\beta$-decay processes. The measured value is $F_A(0)=-1.2723\pm0.0023$, which sets the normalization of the form factor. The axial form factor is also a common target of lattice calculations~\cite{Alexandrou:2017hac,Alexandrou:2020okk,Capitani:2017qpc,Gupta:2017dwj,RQCD:2019jai,Tsuji:2025quu}. For the purpose of extracting the axial radius of the nucleon, the necessary quantity to measure is the slope of the form factor as $t\rightarrow0$, denoted here as $F'_A(t\rightarrow0)$. Under the standard assumption that the form factor follows a dipole form, i.e. $F_A(t)=F_A(0)(1-\frac{t}{M_A^2})^{-2}$, the slope of the form factor is controlled by the dipole mass parameter $M_A$, which can be extracted from the $|t|$-dependence of the scattering cross section (and target-spin asymmetries, as will be discussed later). 

The landscape of existing experimental measurements of $F_A(t)$ includes quasi-elastic neutrino scattering on nuclei, pion electroproduction on protons, and a measurement of $\bar{\nu_\mu}p\rightarrow\mu^+n$ on the organic scintillator of the MINERvA experiment~\cite{Choi:1993PRL_AxialPionElectroprod,Liesenfeld:1999PLB_AxialFromPeepiN,T2K:2020CC0piOC,AguilarArevalo:2010MiniBooNECCQE,MINERvA:2023avz}. Both the neutrino data with nuclear targets and the pion electroproduction data suffer from large epistemic uncertainties and model dependence~\cite{NuSTEC:2017hzk,Haberzettl:2000sm}\footnote{A recent follow-up article from the MINERvA collaboration further highlights this point~\cite{Meyer:2025rzh}.}. The MINERvA data therefore provide the cleanest existing measurement of the axial form factor, and the value of $r_A$ extracted from that data is $0.73\pm0.17$ fm, which is consistent with the charge radius of the proton within the uncertainties. 

The goal of this manuscript is to investigate the feasibility of measuring the CCE process at the EIC. The highly polarized proton beam at the EIC provides the opportunity to gain novel information about the axial form factor via the target-spin asymmetries of the CCE process~\footnote{Throughout this paper, ``target" in the context of the EIC always refers to the proton beam, as is standard parlance in hadronic physics. Target-spin asymmetries thus refer to differences in the process cross section between different proton polarization directions. This is despite the fact that at the EIC there is nominally no fixed target in the traditional sense (see, however, Ref.~\cite{CFNS2025FixedTargetEIC}).}. A polarized proton target is generally infeasible in neutrino scattering experiments due to the large target mass required to acquire adequate statistics for a measurement. Refs.~\cite{Tomalak:2020zlv,Graczyk:2019xwg} calculate the target-spin asymmetries for the crossed process $\nu n\rightarrow l^-p$ for both longitudinal and transverse target polarizations. These target-spin asymmetries are calculated to be very large, the longitudinal asymmetry $A_{UL}$ reaches nearly 100\% at low $Q^2$ and the transverse asymmetry  $A_{UT}$ is greater than 50\% for $Q^2>0.5\mathrm{~GeV}^2$. These asymmetries have never been measured, and it seems unlikely that they can be measured on proton targets at any existing or proposed facility, with the possible exception of the EIC. The $|t|$-dependence of these target-spin asymmetries furthermore encodes information about the $|t|$-dependence of $F_A$, and therefore they can be used to augment the extraction of $r_A$ from scattering cross sections~\cite{Graczyk:2019xwg}. A combined fit of $F_A(t)$ to the target-spin asymmetry and $|t|$-dependence of the cross section of the CCE process in $ep$ scattering would thus maximize the information about the form factor that could be obtained at the EIC.


Speculatively, the CCE process could also be sensitive to beyond-the-Standard-Model (BSM) physics that manifests at energies higher than existing neutrino beam experiments (100 GeV in the target rest frame). BSM interactions mediated by higher-order currents could enhance the CCE cross section relative to the Standard Model prediction while remaining consistent with limits from charged-current inelastic scattering.

\section{Measurement of Charged-Current Elastic Scattering at the EIC} \label{sec:Meas}

Compared to existing and proposed measurements of the axial form factors via (anti)neutrino scattering, performing the measurement with an electron beam provides some obvious benefits. First, the beam energy is precisely known. Electron beams can also generally attain much greater intensities than neutrino beams. Another useful feature is that electron beams often have longitudinal polarization. Since the CCE process is 100\% parity-violating, only the electron polarization state where the electron spin is anti-aligned with its momentum can participate in the signal reaction. Selecting the aligned electron polarization state can therefore provide a data-driven estimate of the background. These aspects, among others, have also been discussed in a recent proposal to measure the axial form factor in electron scattering at Jefferson Lab~\cite{JLab-PR12-25-009}. 

There are also some obvious challenges in measuring the CCE process with electron scattering. The first and most daunting challenge is the background, which is discussed in detail in Sec.~\ref{sec:bkg}. Furthermore, the fully neutral final state of $\nu_e+n$ poses an experimental challenge, as neutrinos realistically cannot be measured and measurement of neutron energies and angles is generally difficult. One of the most powerful techniques for suppressing background in elastic scattering is to exploit the kinematic exclusivity, which is difficult with poor resolution on energy and scattering angle.

\begin{figure}[t]
    \centering
    \qquad 
    \includegraphics[width=1.0\linewidth]{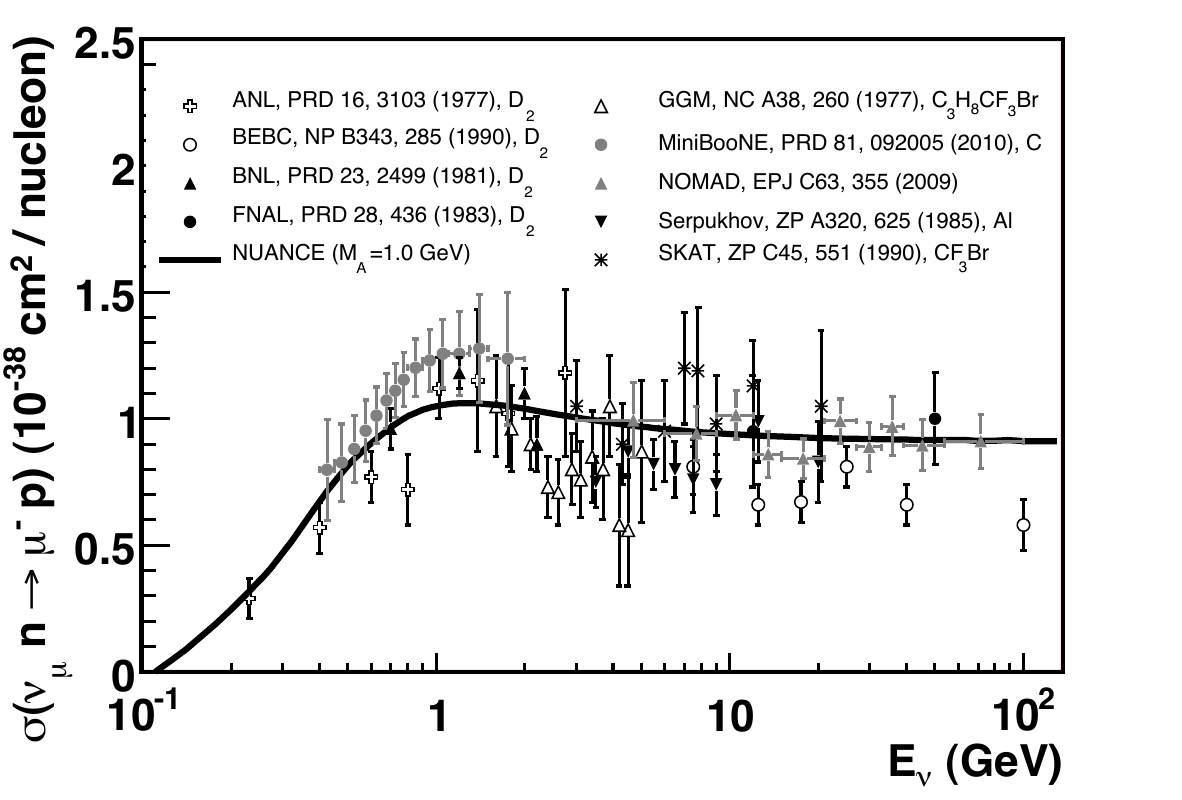}
    \caption{Compilation of data for the reaction $\nu_{\mu} n\rightarrow \mu^- p$ reproduced from Ref.~\cite{Formaggio:2012cpf}. By crossing symmetry, this cross section should be the same as $e^-p\rightarrow\nu_en$, neglecting subleading effects such as lepton masses and higher-order corrections.
}
    \label{fig:NuData}
\end{figure}

The EIC, at its design luminosity, will deliver around 100 fb$^{-1}$ per year in the 10 GeV electron on 275 GeV proton (henceforth denoted 10x275) configuration~\cite{AbdulKhalek:2021gbh}. The corresponding energy of the electron in the frame where the proton is at rest for the 10x275 GeV configuration is 5.9 TeV. To estimate the value of the CCE cross section at the EIC we perform a linear fit of the world data on $\nu_\mu n\rightarrow\mu^-p$ shown in Fig.~\ref{fig:NuData} for $E_{\nu}>3$ GeV. Utilizing crossing symmetry and neglecting differences between muons and electrons, we arrive at an expected cross section for CCE at $E_{\mathrm{lab}}=5.9$ TeV of roughly $7\times10^{-39}~\mathrm{cm}^2$, i.e. 7 femtobarns. Since this cross section corresponds only to left-handed leptons and the EIC beam will be an equal mix of left- and right-handed electrons, the expected event rate for the EIC at maximum luminosity is around 350 events per year. CCE is therefore a prime example of a process that would have been impossible to measure at HERA but is potentially within reach thanks to the high luminosity of the EIC.

\begin{figure}[t]
    \centering
    \qquad 
    \includegraphics[width=0.9\linewidth]{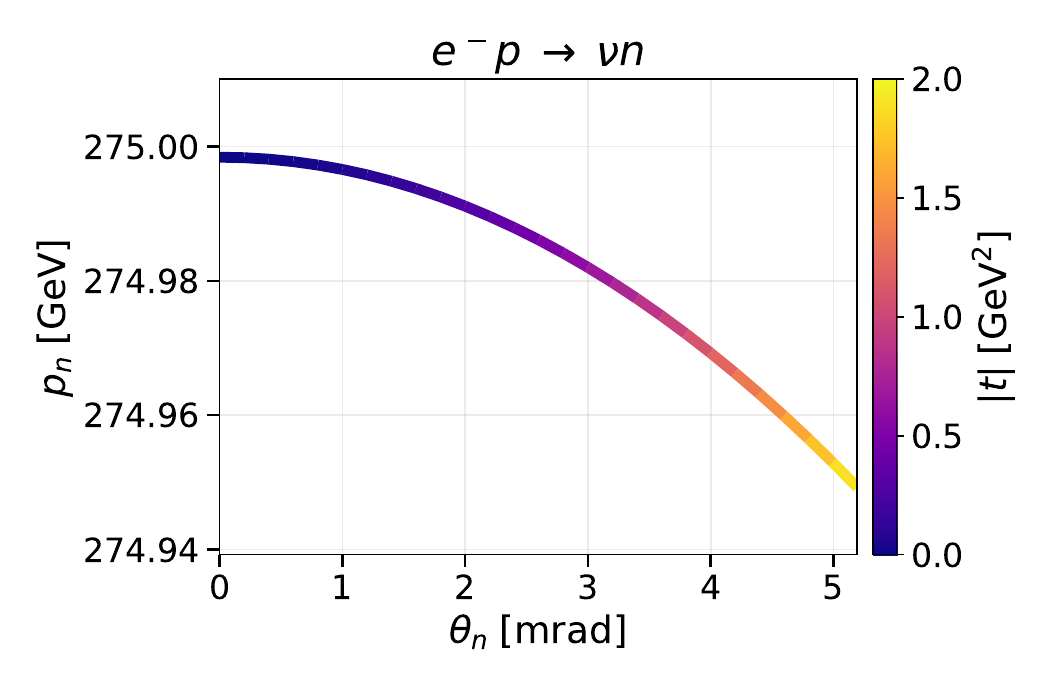}
    \caption{Plot of neutron momentum vs. lab-frame scattering angle for the CCE reaction. Note the small change in neutron momentum as a function of $|t|$ on the $y$-axis.}
    \label{fig:EvsTheta}
\end{figure}

Currently, the EIC project includes one detector, the Electron-Proton/Ion Collider (ePIC) experiment, with the possibility of a second detector. The benefit of searching for the CCE process at the EIC is that the produced neutrons will be at very high energies, making them easier to detect and distinguish from non-physics backgrounds. The detector subsystem of ePIC that will measure the elastically scattered neutron is the Zero-Degree Calorimeter (ZDC), which employs an SiPM-on-tile design~\cite{Blazey:2009zz,Simon:2010hf,Liu:2015cpe,Magnan:2017exp,Sefkow:2018rhp}. This design is expected to achieve excellent resolutions on both neutron energies and neutron positions~\cite{Milton:2024bqv,Zhang:2025ijg}. The anticipated energy resolution using a graph neural network reconstruction technique is $\frac{\sigma_E}{E}=29\%/\sqrt{E}\oplus2\%$, and the position resolution for neutrons near the beam energy is on the order of 0.04 mrad. As can be seen from Fig.~\ref{fig:EvsTheta}, for $|t|<2\mathrm{~GeV^2}$ the final-state neutron has a momentum indistinguishable from the beam momentum. Using an angular resolution of 0.04 mrad, the resolution on $|t|$ for the CCE process is around 5\%. However, the angular divergence of the proton beam in the highest luminosity configuration actually dominates the resolution, providing a contribution to the angular resolution of 0.119 mrad~\cite{EIC_MasterParameter_Table_2023}. 


\begin{figure}[h]
    \centering
    \qquad 
    \includegraphics[width=0.9\linewidth]{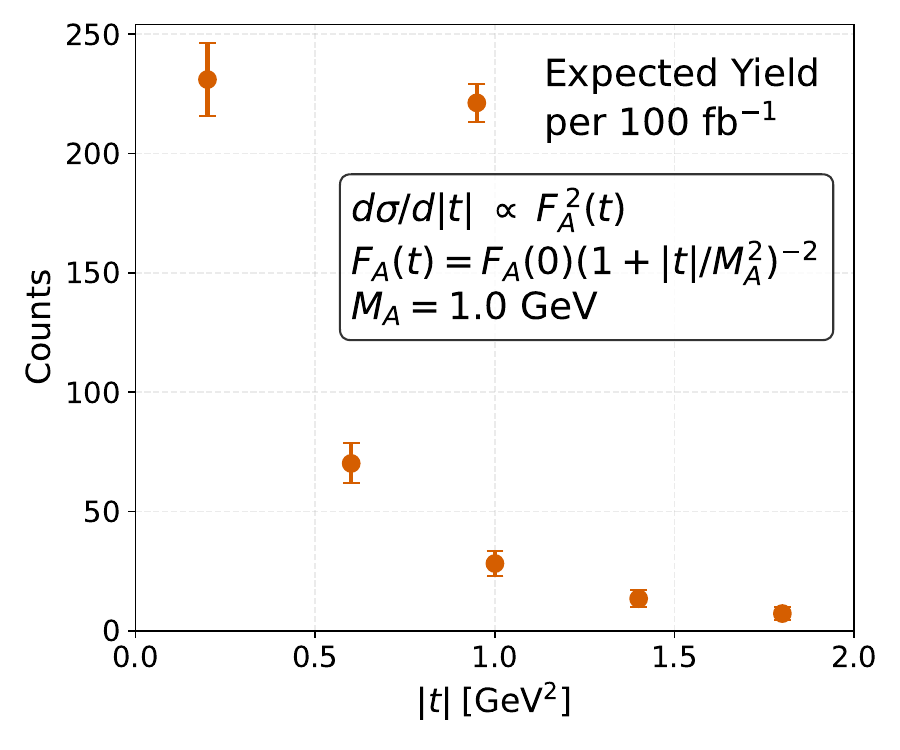}
    \caption{Expected event yield in 100 fb$^{-1}$ of integrated luminosity for the EIC in the 10x275 GeV configuration. The bin width in $|t|$ is 0.4 GeV$^2$.}
    \label{fig:Yield}
\end{figure}

\section{Background}
\label{sec:bkg}

Since the CCE process is reconstructed from a single neutron, processes which produce high-energy neutrons serve as a background. To understand the background from inclusive leading neutron photoproduction, $\gamma+p\rightarrow n+X$, the ZEUS data at $\sqrt{s}=300$ and $\sqrt{s}=318$ GeV provide a useful starting point~\cite{ZEUS:2002gig,ZEUS:2007knd}. ZEUS provides photoproduction cross sections as a function of $x_L$, defined as
\begin{align}
    x_L=\frac{N\cdot k}{P\cdot k}\approx\frac{E_N}{E_P},
\end{align}
where $k$ and $P$ are the initial electron and proton beam four-vectors, respectively, $N$ is the produced neutron four-vector, and $E_P$ is the incident proton beam energy. The ZEUS data at $E_p=820$ GeV suggest that the total $\gamma p$ cross section for this process in the region of $x_L>0.88$ and $\langle W\rangle=\mathrm{209~GeV}$ is around 0.6 $\mu$b for $Q^2<0.02$ GeV$^2$. Given that the fraction of $ep$ events with a leading neutron seems to be approximately independent of $Q^2$ and that the leading neutron structure functions do not vary significantly in this region of $Q^2$~\cite{ZEUS:2002gig}, we can use the equivalent photon approximation to simply scale the cross section for $Q^2<0.02$ GeV$^2$ and $0.345 < y < 0.636$ to the expected total cross section for $Q^2<1$ GeV$^2$ based on the photon flux. The choice of 1 GeV$^2$ corresponds to the point at which the ePIC central detectors should be able to efficiently veto the scattered electron. This scaling methodology is described in the Appendix~\ref{app:flux}, and the total $ep$ cross section for $Q^2<1$ GeV$^2$ and $0<y<1$ is expected to be around $0.3$ $\mu$b for $x_L>0.88$, $Q^2 < 1$ GeV$^2$, and $0<y<1$.

The result is that the signal-to-background for CCE events at the EIC is expected to be around $10^{-7}$ to $10^{-8}$. Unfortunately, the shape of the $|t|$-distribution of the background will likely be similar to that of the signal, although it can be measured in several ways. The shape of the background can be measured using the right-handed electron beam bunches as a control sample or by using events where a scattered electron is detected in the Low-$Q^2$ tagger. To have a measurement of $5\sigma$ statistical significance for the expected 350 events in one year of running, only around 5000 background events can survive the vetoing process, assuming a perfect knowledge of the background distribution. This means that the required background rejection factor necessary for 100 fb$^{-1}$ is on the order of $10^7$, decreasing to $10^6$ if 1 ab$^{-1}$ is collected. Rejection of background at these levels is common in the realm of rare decay experiments but is certainly a unique challenge in a collider environment. At colliders, efficiency for reconstruction of particles is typically valued over efficiency for vetoing particles, and the latter is in general poorly studied for the planned EIC detector subsystems. For most exclusive channels, which would benefit the most from high veto efficiency, the kinematic constraints imposed by exclusivity are enough to reduce backgrounds to the required levels. Since the effectiveness of the ePIC subdetectors for vetoing has not been studied in detail (with the exception of Ref.~\cite{Aschenauer:2025mku}), a quantitative estimate of how much the background can be reduced is at present beyond the scope of this work. The remainder of this section will be dedicated to some strategies that could be used to reduce the background for the CCE process.

While the angular resolution from the beam divergence dominates the measurement of $|t|$, the energy resolution of the ZDC still plays a vital role in the measurement, namely for vetoing background from lower $x_L$ leading neutron events. No inelastic process can produce neutrons as close to the beam energy as the CCE process. Therefore the ability to reject neutrons with incident energies below the endpoint will drastically reduce the background. At the endpoint energy, the expected energy resolution of the ZDC is 2.6\%, i.e. 7.2 GeV. Crucially, this energy resolution does not depend on $\theta_n$ for the range of scattering angles corresponding to $|t| < 2~\mathrm{GeV}^2$. It can be seen from Fig.~\ref{fig:EvsTheta} that all CCE data reside at $x_L=1$, to a good approximation. A cut on neutron energies measured more than $2\sigma_E$ away from the endpoint $x_L=1$ would remove $x_L<0.948$. In Ref.~\cite{ZEUS:2007knd}, ZEUS measures that in a bin of $0.95 < x_L < 1.00$, the leading neutron cross section is around 7 times smaller than the cross section in the bin from $0.90 < x_L < 0.95$. Since we are concerned only with the kinematic endpoint of $x_L\approx1$, we can gain around a factor of 10 suppression compared to the $x_L>0.88$ cross section by selecting neutrons within $2\sigma_E$ of the beam energy. A better energy resolution on the ZDC, albeit hard to imagine since the assumed resolution is already better than any previous hadronic calorimeter at a collider, would help by reducing the region of neutron energies that need to be accepted.

The most straightforward route to veto non-CCE events is to veto any event containing a scattered electron, including those associated with photoproduction events. Electroproduction of leading neutrons for $Q^2>1$ GeV$^2$ should be able to be efficiently vetoed by measuring the scattered electron in the ePIC central detector. Based on their anticipated efficiencies, the lead tungstate backward calorimeter~\cite{Houzvicka:2022pny,Philip:2024opz,Mkrtchyan:2025euu}, proximity-focusing RICH detector~\cite{Bhattacharya:2023ddr,Chatterjee:2024zrn}, and backward tracker~\cite{Sidoretti:2025tja,Sidoretti:2025xeh,Gonella:2024cxk,Li:2023hjr} should combine to a background rejection factor similar to the one required for $Q^2>1$ GeV$^2$. Some of the $Q^2<1$ GeV$^2$ phase space will be covered by the far-backward detectors~\cite{Gardner:2023lly}. In principle, the ePIC low-$Q^2$ tagger has acceptance for $10^{-7}<Q^2<0.01$ GeV$^2$, although this detector cannot be used as a reliable veto since it has a relatively low acceptance and a high occupancy per bunch crossing from Bethe-Heitler events~\cite{Gardner:2023lly}.

Vetoing on the scattered electron can remove the electroproduction reaction, but the photoproduction cross section dominates the rate of leading neutron events. In photoproduction of neutrons ($\gamma+p\rightarrow n + X$) at larger $|t|$, where the $p_T$ of the neutron is appreciable, the non-neutron hadronic final state (denoted here simply HFS) $X$ must balance the neutron $p_T$, enabling it to be more efficiently vetoed in the central and far-forward detectors. Therefore, it could be that an observation of this process above background is possible only at larger values of $|t|$, e.g. $|t|\gtrsim1~\mathrm{GeV^2}$, where the missing transverse momentum is large. The total longitudinal momentum of the produced hadrons in the proton-going direction can be at maximum $(1-x_L)E_p$. For a 2$\sigma$ energy cut around the endpoint of $E_n=E_p$, assuming the ePIC ZDC energy resolution given above, the maximum $p_z$ available for the hadronic final state is therefore 14.3 GeV, while the maximum $p_T$ for the HFS is the $p_T$ of the final-state neutron. In the high divergence configuration, at $|t|=1~\mathrm{GeV^2}$ the resolution on the $p_T$ of the neutron is about 45 MeV. At larger values of the neutron $p_T$, vetoing one or more charged particles balancing that $p_T$ becomes easier. The angular and energy resolutions on the neutron constrain the hadrons produced in background events to be primarily at less than roughly 5\textdegree~($\eta\geq3$) with respect to the proton beam direction. Furthermore, for each neutron there will be an expected location where the balancing hadron or electron for background events should appear in the detector, adding additional information that can be used in the veto procedure. Unfortunately, in this case missing mass does not provide a very discriminating quantity to reject background because of the resolution. 

\begin{figure}[h]
    \centering
    \qquad 
    \includegraphics[width=0.9\linewidth]{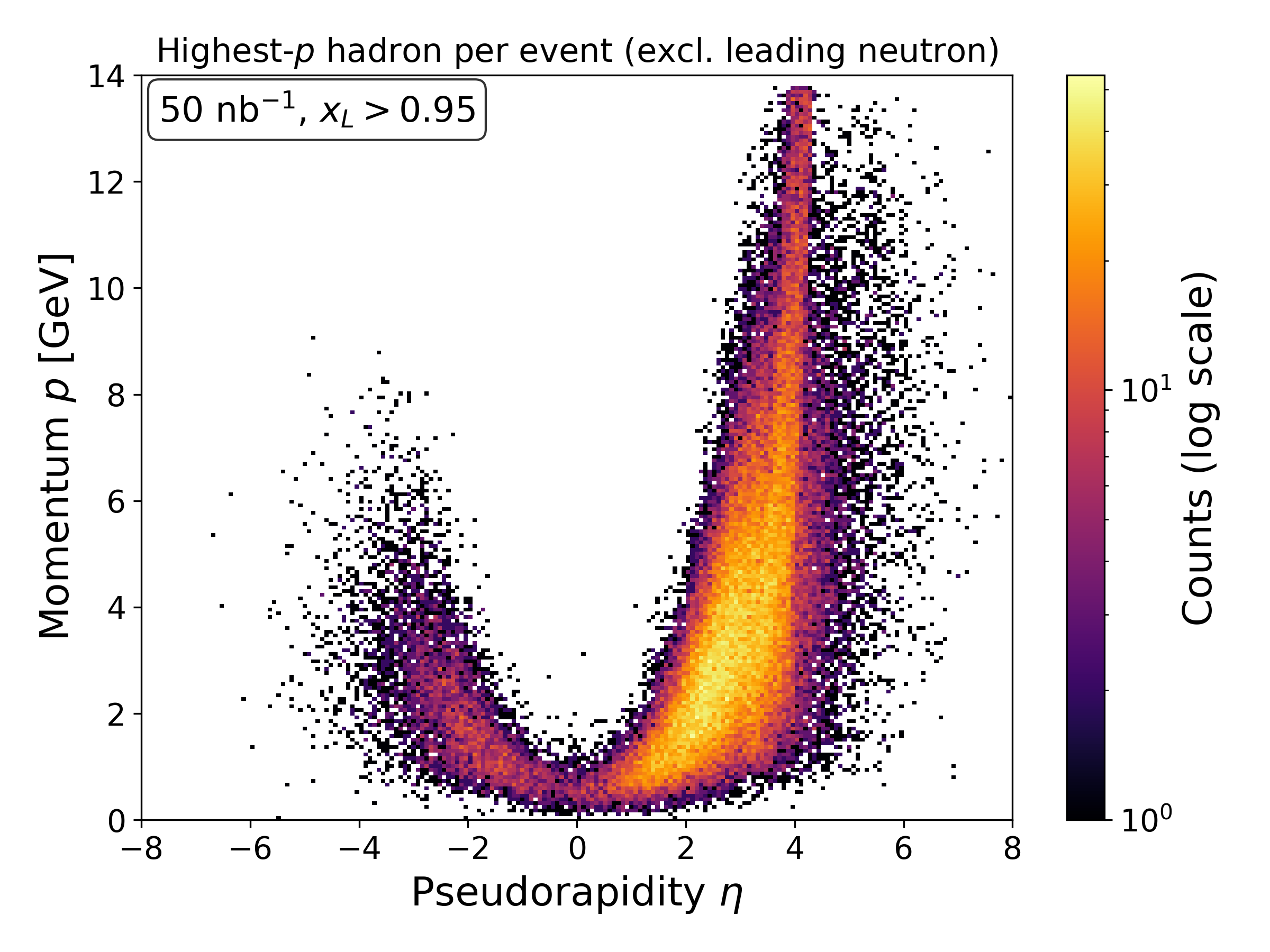}
    \caption{Momentum vs. pseudorapidity for the highest-momentum charged particle in an event with an $x_L>0.95$ leading neutron as simulated by Pythia6~\cite{Sjostrand:2006za}.}
    \label{fig:BackgroundPvEta}
\end{figure}

It is important to note that all background reactions must have a charged hadron in the final state while the signal reaction does not. Vetoing these charged hadrons provides essentially the only hope to reduce the background to a manageable level. Charged hadrons produced near the proton beam ($\eta>4.5$) will be swept out of the beampipe at the B0 dipole and at that point can be vetoed to reduce the background. To get an idea of the distributions of hadrons in leading-neutron events, we use Pythia6~\cite{Sjostrand:2006za}. The events were generated with no cuts on kinematic variables, i.e. including photoproduction, and the distribution of charged hadron momenta and pseudorapidity are shown in Fig.~\ref{fig:BackgroundPvEta}. The band at large $\eta$ and momentum corresponds to $\Delta^+$ baryon production and decay to $\pi^+n$. It can be seen that vetoing charged hadrons in the range of $\eta>3.5$ is crucial to veto events with high-energy neutrons produced from low-mass baryon resonances.

\section{Veto System}
\label{sec:bkg}

With an understanding of the relevant backgrounds, we now turn to what can be done about them. The current design of ePIC has gaps in the forward acceptance that preclude a measurement of the CCE process. However, a dedicated system to veto hadronic particles produced in the forward direction with high efficiency could be envisioned for a second detector at the EIC or an upgrade of ePIC.

\begin{figure}[h]
    \centering
    \qquad 
    \includegraphics[width=0.9\linewidth]{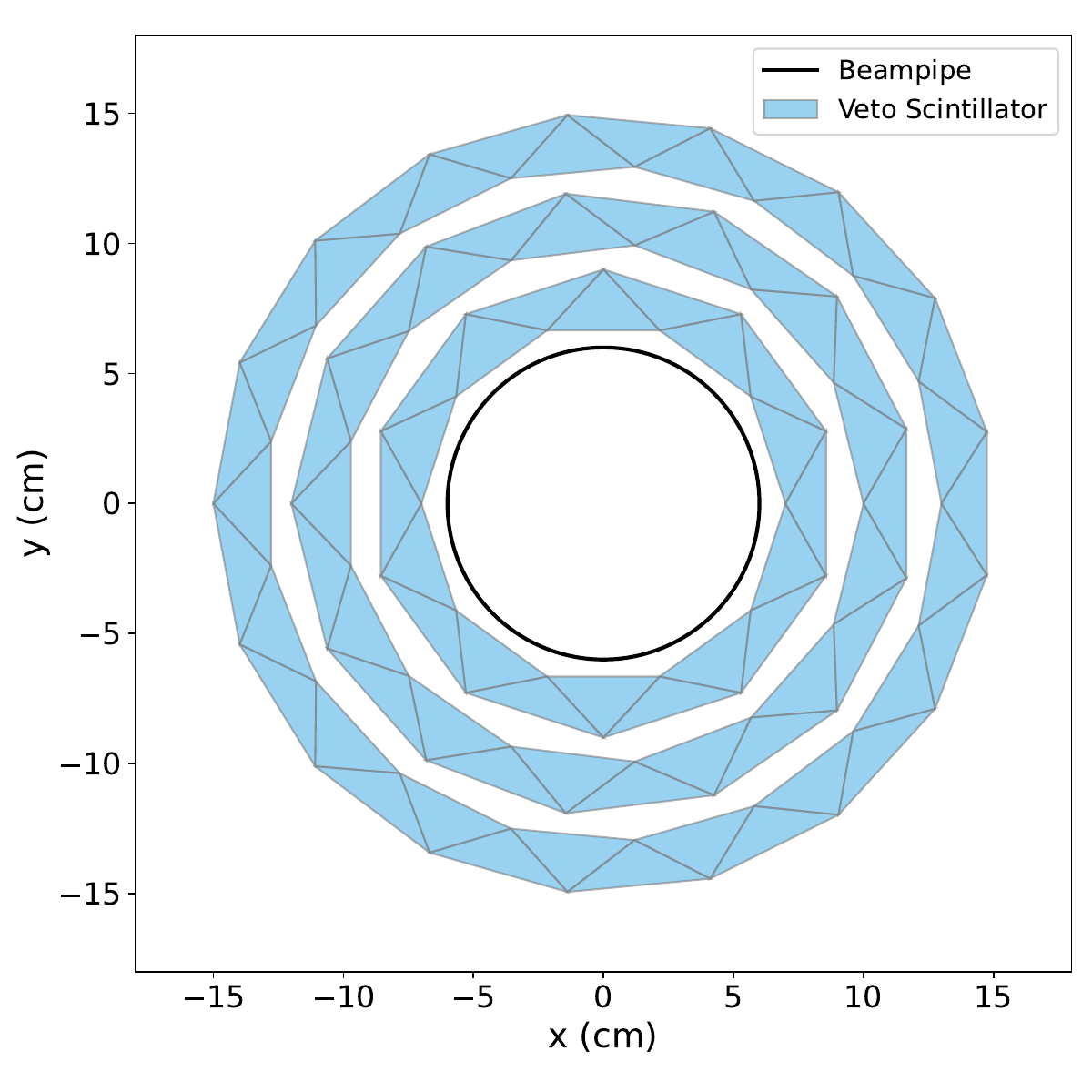}
    \caption{Sketch of a veto detector that would surround the beampipe to veto any charged particle.}
    \label{fig:vetodet}
\end{figure}

As an example of a successful veto system, layered plastic scintillator detectors with single-layer efficiencies of 99.8\% for single charged particles were built and operated for the NA62 experiment~\cite{Ambrosino:2015ruf}. Three layers of similar detectors arranged continuously around the beampipe in a configuration similar to the one sketched in Fig.~\ref{fig:vetodet} could provide the necessary veto efficiency of $\approx10^6$. The shape of the scintillators will necessarily conform to the final shape of the beampipe, which has a complicated geometry in the far-forward region~\cite{Aschenauer:2025mku}.

Veto systems based on inexpensive plastic scintillators were used extensively in diffractive measurements at HERA to improve the selection of diffractive events by vetoing hadronic activity between the proton remnant and the diffractively produced final state. Using these scintillators, H1 gained the ability to veto particles produced in the region of $-3.5<\eta<7$, greatly improving the acceptance and purity of diffractive events. Such a system of scintillators at the EIC would serve a similar purpose, providing benefits particularly for inclusive and semi-inclusive diffractive measurements, in addition to enabling the CCE measurement.

A unique challenge for a scintillator-based charged particle veto system at the EIC is that the instantaneous rate of hits per channel can be large, depending on the granularity that is chosen. The expected averaged event rate at the EIC at top luminosity is around 500 kHz, and a large fraction of those events will produce hits in the forward region. A non-negligible hit rate from beam-gas events and synchrotron radiation can also be expected. The bunches at the EIC are spaced by roughly 10 nanoseconds. Two closely timed hits, e.g. from two consecutive bunch crossings, in the scintillator can result in the later hit not being properly vetoed if the response time of the scintillator and photosensor is too long. From that perspective, a photosensor with a fast rise- and fall-time such as an MCP-PMT may be desired to avoid hits shortly after one another piling up and defeating the veto. The decay time of plastic scintillator, for example formulas based on PPO fluorescence with POPOP wavelength shifter, is commonly as short at 2 ns, making it ideal for this use case~\cite{Eljen_EJ200_204_208_212}. The expected radiation dose in the EIC far-forward region is approximately $3\times10^{11}$ 1 MeV neutron equivalent of non-ionizing radiation and 100 krad of ionizing radiation in 100 fb$^{-1}$. Plastic scintillators have been shown to retain their performance at significantly higher radiation doses~\cite{Kharzheev:2019tfk}. Finally, plastic scintillator can be extruded or molded into awkward shapes, as will likely be necessary for hermetic coverage around the beampipe. 

\section{Projections}
\label{sec:results}

\begin{figure*}[htbp]
    \centering
    \qquad 
    \includegraphics[width=0.99\linewidth]{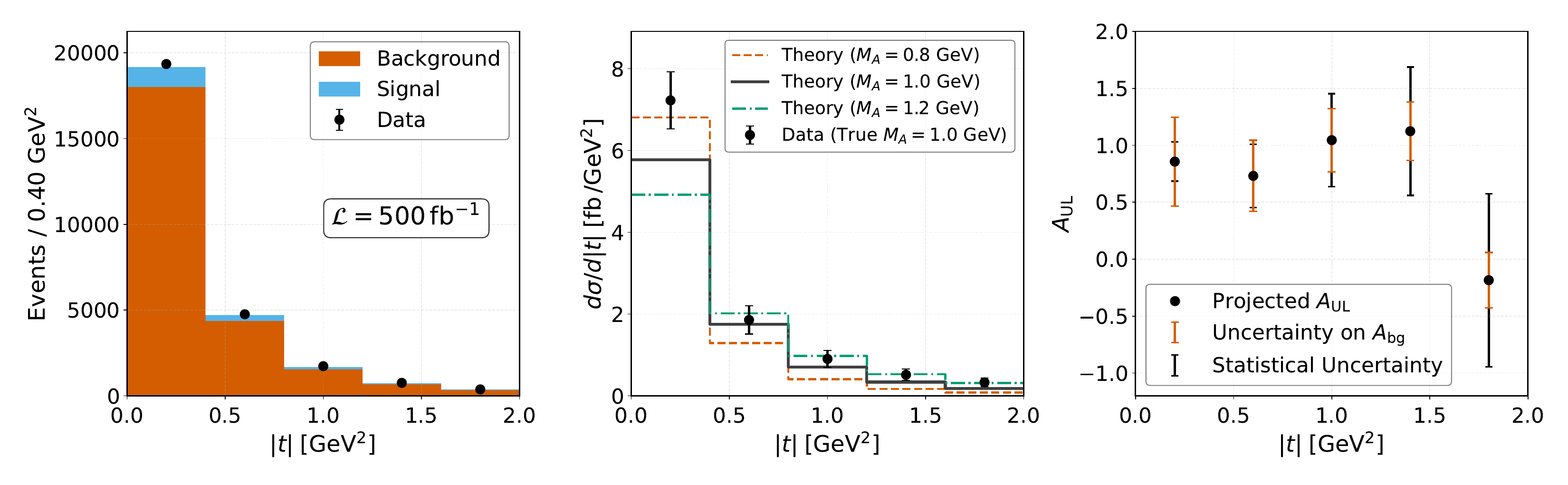}
    \caption{\textbf{Left:} Projected yields as a function of $|t|$ for 500 fb$^{-1}$. \textbf{Center:} Background-subtracted CCE cross section. \textbf{Right:} Projected target-single-spin asymmetry including uncertainty on the background asymmetry $A_{bg}$. See text for more details.}
    \label{fig:results}
\end{figure*}
Assuming a veto system capable of reducing the number of background events to 5000 per 100 fb$^{-1}$, we make projections for the measurement of the CCE cross section and spin asymmetries in Fig.~\ref{fig:results}. The shape of the background is assumed to have the same functional form as the signal shown in Fig.~\ref{fig:Yield}, but with a different dipole mass of 0.8 GeV to avoid bias from having exactly the same background and signal distribution shapes. An integrated luminosity of 500 fb$^{-1}$, corresponding to 5 years of running at the design luminosity, is used for these projections. The first two panels of Fig.~\ref{fig:results} show the total expected yield including background (left) and expected cross section measurement (center) in 500 fb$^{-1}$ of integrated luminosity. In the background-subtracted cross section projection, the uncertainty is only statistical, taken from the total observed yield in each $|t|$ bin, $\delta N=\sqrt{N_{\rm obs}}$, such that it includes the Poisson fluctuations from both signal and background events. The true data are generated at $M_A=1.0$ GeV, and cross section curves corresponding to $M_A$ values of 0.8 GeV and 1.2 GeV are overlaid. A precision on $M_A$ of around 20\% could be expected from a dipole fit to these datapoints. In the dipole ansatz, $r_A$ can be expressed solely in terms of $M_A$ as $r_A^2=12M_A^{-2}$, such that a 20\% uncertainty on $M_A$ would correspond to a 20\% uncertainty on $r_A$ prior to inclusion of target-spin asymmetry data. This precision would be on par with the MINERvA result if it can be realized experimentally, although an excellent understanding of the background and large integrated luminosity would be required to make it a reality.

The right panel of Fig.~\ref{fig:results} show projections for the $|t|$-dependent longitudinal target-spin asymmetry. The longitudinal target-spin asymmetry is defined as
\begin{equation}
A_{UL} \equiv \frac{1}{P_p}\,\frac{N^{\leftarrow}-N^{\rightarrow}}{N^{\leftarrow}+N^{\rightarrow}},
\end{equation}
where $N^{\leftarrow(\rightarrow)}$ are the event yields for the two proton-helicity states and $P_p$ is the average value of the proton longitudinal polarization. The projections for $A_{UL}$ are shown in the right panel of Fig.~\ref{fig:results}. A value $P_p=0.70\pm0.01$ is assumed for the proton beam polarization, consistent with the EIC expectations~\cite{Huang:2025gqx,Rathmann:2025jgp,Atoian:2025dib}. The proton beam polarization can be either longitudinal or transverse at the interaction point. The degree of proton polarization is monitored via dedicated hadron polarimeters that make use of the left-right asymmetry of the beam scattering on a hydrogen gas jet or carbon foil target~\cite{AbdulKhalek:2021gbh,Nunes:2022cmc}. Propagating the expected $1\%$ proton polarization uncertainty gives a fractional uncertainty $\delta A_{UL}/A_{UL}\simeq \delta P_p/P_p \approx 1\%$, significantly smaller than the statistical uncertainty for all $|t|$ bins. Thus the beam polarization uncertainty is negligible compared to the projected statistical uncertainties, and in fact a larger polarization uncertainty would be tolerable for this measurement. 

The longitudinal target-spin asymmetry of the background process $e\vec{p}\rightarrow n+X$ is currently unknown, but its magnitude will impact how well the CCE asymmetry can be extracted. To account for the target-spin asymmetry of the background that will need to be subtracted, we assume the background has an asymmetry of $0.05\pm0.025$ and propagate this uncertainty into the final uncertainty on the measured CCE asymmetry. This source of uncertainty can be reduced significantly using even low-luminosity data from the early runs of the EIC to measure the $e\vec{p}\rightarrow n+X$ asymmetry. The datapoints are scattered around a nominal asymmetry of $A_{UL}=1$ according to their total uncertainties to mimic a real measurement. The results for a transverse-spin asymmetry $A_{UT}$ would be the similar to those shown in Fig.~\ref{fig:results}, barring a very large $A_{UT}$ for the background process $e\vec{p}\rightarrow n+X$. Both transverse and longitudinal proton spin asymmetries can be utilized to constrain $F_A(t)$, provided enough integrated luminosity is available.



The results of Ref.~\cite{Graczyk:2019xwg} (in particular Fig. 4 of Ref.~\cite{Graczyk:2019xwg}) show the strong momentum-transfer dependence of the longitudinal target-spin asymmetry on the value of the dipole mass $M_A$. These results should be taken with a grain of salt when projecting to the EIC kinematics since the correspond to an incoming neutrino energy of only 1 GeV. Nevertheless, a combined fit of $M_A$ to the $|t|$-dependent spin asymmetries and cross section will provide the best constraint, and is a unique possibility offered by the EIC. Finally, the measurement of these spin asymmetries will be valuable generically to cross-check existing results and theoretical understanding with a novel technique\footnote{One is reminded of the proton elastic form factor puzzle~\cite{JeffersonLabHallA:1999epl}, where the inclusion of the polarization transfer technique for measuring the nucleon $G_E/G_M$ form factor ratio revealed a large discrepancy with the standard Rosenbluth method.}.


\section{Remarks on Other Charged-Current Cross Sections}
\label{sec:remarks}

It can be seen from Fig.~\ref{fig:NuData} that the CCE total cross section remains approximately constant with $\sqrt{s}$ due to competition between the nucleon form factors and increased effective flux of $W^-$. On the other hand, the inclusive charged-current cross section increases linearly with $E_\nu$ due to increasing contribution of charged-current DIS, which dominates above $E_\nu\approx20$ GeV. The phase space in $\gamma p$ center-of-mass energy $W$ for hard exclusive processes such as deeply virtual Compton scattering (DVCS) and deeply virtual meson production (DVMP) are much larger than for elastic scattering, and the DVCS and DVMP cross sections are known to grow with $W$~\cite{H1:2005gdw,ZEUS:2008hcd,H1:2009cml,Kumericki:2009uq}. One may expect then that at EIC energies, the cross sections for charged-current exclusive processes such as CCDVCS ($e^-p\rightarrow\nu_e n\gamma$) or CCDVMP (e.g. $e^-p\rightarrow\nu_e p\pi^-$) could be larger than the CCE cross section and counterintuitively may be more experimentally favorable. This indeed appears to be the case, as the cross sections for $\pi^-$ and $D_s^-$ meson production at the EIC were calculated using GPD models to be reasonably large compared to the CCE cross section in Refs.~\cite{Siddikov:2019ahb,Pire:2021dad}. While the combination of relatively low cross section and unfavorable experimental branching fractions may make $D_s^-$ challenging, the CCDVMP $e^-p\rightarrow\nu_e p\pi^-$ reaction could be a nice experimental target for the EIC since the $\pi^-$ is measured directly and not through a decay. The all-charged hadronic final-state of $p\pi^-$ will also have excellent resolution on missing mass and missing transverse momentum since the particles can both be measured in trackers. The charged-current meson production process is also expected to have large spin asymmetries that can be exploited similarly to CCE scattering~\cite{Pire:2021dad,Graczyk:2021oyl,Graczyk:2023lrm}. 

\section{Conclusion}

The charged-current elastic scattering process provides a novel means to study the axial structure of the nucleon and potentially some presently unexplored corners of BSM physics. The recent MINERvA analyses of Refs.~\cite{MINERvA:2023avz,Meyer:2025rzh} have demonstrated the power of utilizing a single nucleon target to accurately constrain the nucleon axial structure. From this perspective, the EIC is well-positioned to contribute to this program via measurement of the CCE process with a polarized proton beam. With its high luminosity and polarized beams of electrons and protons, the EIC is, in principle, an excellent tool for studying this reaction. The cross section for CCE scattering at the EIC is expected to be around 3.5 fb assuming an equal mix of left- and right-handed electrons in the electron beam. Target-spin asymmetries, which can be measured for the first time at the EIC, are predicted to be on the order of 100\%, with their magnitudes and $|t|$-dependencies providing additional sensitivity to the axial form factor. 

The presence of a dominating background from inclusive neutron photoproduction makes it highly unlikely this process can be measured at the EIC without dedicated instrumentation. If one is optimistic, a dedicated hermetic veto detector may be able to deliver the necessary background suppression, depending on the final implementation of the far-forward region and the level of machine backgrounds. Such a veto system could be incorporated into the far-forward region of a second EIC detector. Even with such a veto system, it will be challenging for the EIC to deliver a competitive measurement of the axial form-factor due to large statistical uncertainties. Leveraging the full information encoded in the target-spin asymmetries and cross section will provide the best constraint on the nucleon axial charge radius $r_A$.
\section*{Acknowledgments}
 I would like to acknowledge Tim Hobbs, Garth Huber, Sylvester Joosten, Jihee Kim, Zein-Eddine Meziani, Noémie Pilleux, and Maria Żurek for useful discussions and comments. I thank Bernard Pire for bringing Refs.~\cite{Siddikov:2019ahb,Pire:2021dad} to my attention. I also thank Lindsay DeWitt for editing this manuscript. This work was supported by the U.S. Department of Energy under Contract No. DE-AC02-06CH11357.

\appendix
\section*{Appendix: Details of the Photon Flux}
\label{app:flux}
\newenvironment{tightdisplay}{\begingroup
\setlength\abovedisplayskip{1.5pt}%
\setlength\belowdisplayskip{1.5pt}%
\setlength\abovedisplayshortskip{1.5pt}%
\setlength\belowdisplayshortskip{1.5pt}%
}{\endgroup}

This section provides more detail on how the rates for leading neutron production were estimated based on the HERA data of Refs.~\cite{ZEUS:2002gig,ZEUS:2007knd}. The cross sections reported by ZEUS are $\gamma p$ cross sections and therefore must have the photon flux included to obtain an $ep$ cross section. The photon flux from the
electron as a function of $Q^2$ and $y$ is
\begin{align}
\frac{\mathrm d^2 N_\gamma}{\mathrm d y\,\mathrm d Q^{2}}
&=\frac{\alpha}{2\pi}\,\frac{1+(1-y)^2}{y}\,\frac{1}{Q^{2}}
-\frac{\alpha}{2\pi}\,\frac{2 m_e^{2}\,y}{Q^{4}}
\notag\\
&=\frac{\alpha}{2\pi}\,\frac{1+(1-y)^2}{y}\,\frac{1}{Q^{2}}
\!\left(1-\frac{Q_{\min}^2(y)}{Q^{2}}\right),
\tag{A1}
\end{align}
with
\begin{equation}
Q_{\min}^2(y)=\frac{m_e^2\,y^2}{1-y}.
\tag{A2}
\end{equation}
This $Q_{\min}^2$ comes from the electron mass and the kinematics of emitting a quasi-real
photon at inelasticity $y$.

Assuming the $\gamma p\rightarrow n+X$ subprocess is approximately independent of $Q^2$ and $y$ in the quasi-real
region, the $Q^2$–integrated $ep$ cross section up to $Q_{\max}^2$ at fixed $y$ is
\begin{tightdisplay}
\begin{multline}
\sigma_{ep}(Q^{2}<Q_{\max}^{2};y)\simeq
K(y)\!\int_{Q_{\min}^{2}(y)}^{Q_{\max}^{2}}
\frac{\mathrm d Q^{2}}{Q^{2}}
\left(1-\frac{Q_{\min}^2(y)}{Q^{2}}\right)
\\
=\,K(y)\!\left[
\ln\!\frac{Q_{\max}^{2}}{Q_{\min}^{2}(y)}
-1+\frac{Q_{\min}^{2}(y)}{Q_{\max}^{2}}
\right],
\tag{A3}
\end{multline}
\end{tightdisplay}

where
\begin{tightdisplay}
\begin{equation}
K(y)=\frac{\alpha}{2\pi}\,\frac{1+(1-y)^2}{y},
\tag{A4}
\end{equation}
\end{tightdisplay}
and the expression is taken to be zero when $Q_{\min}^2(y)>Q_{\max}^2$. To compare two $y$ ranges, we integrate over $y$ between lower and upper bounds $y_a$ and
$y_b$:
\begin{tightdisplay}
\begin{multline}
\sigma_{ep}(Q_{\max}^{2};y_a,y_b)
=\\
\int_{y_a}^{y_b}\!dy\,K(y)\,
\Bigl[
\ln\!\frac{Q_{\max}^{2}}{Q_{\min}^{2}(y)}
-1+\frac{Q_{\min}^{2}(y)}{Q_{\max}^{2}}
\Bigr]
\\
\times \Theta\!\bigl(Q_{\max}^{2}-Q_{\min}^{2}(y)\bigr).
\tag{A5}
\end{multline}
\end{tightdisplay}
\hspace{0.08cm}
It is convenient to write the $Q_{\max}^{2}=1$ versus $Q_{\max}^{2}=0.02~\mathrm{GeV}^{2}$ ratio
over the \emph{same} $y$ interval as:
\begin{tightdisplay}
\begin{align}
\mathcal R(y_a,y_b) &= \frac{N(y_a,y_b)}{D(y_a,y_b)}. \tag{A6}
\end{align}

\begin{equation}\tag{A7}
\begin{aligned}
N(y_a,y_b) &= \int_{y_a}^{y_b}\!dy\,K(y)\,
  \Bigl[\ln\!\tfrac{1}{Q_{\min}^{2}(y)}-1+Q_{\min}^{2}(y)\Bigr] \\
&\quad \times \Theta\!\bigl(1-Q_{\min}^{2}(y)\bigr)
\end{aligned}
\end{equation}

\begin{equation}\tag{A8}
\begin{aligned}
D(y_a,y_b) &= \int_{y_a}^{y_b}\!dy\,K(y)\,
  \Bigl[\ln\!\tfrac{0.02}{Q_{\min}^{2}(y)}-1+\tfrac{Q_{\min}^{2}(y)}{0.02}\Bigr] \\
&\quad \times \Theta\!\bigl(0.02-Q_{\min}^{2}(y)\bigr)
\end{aligned}
\end{equation}
\hspace{0.05cm}

\end{tightdisplay}
The result is that expanding the integration range from $Q^2<0.02~\mathrm{GeV^2}$ and $0.435<y<0.636$ to $Q^2<1~\mathrm{GeV^2}$ and $0.0001<y<0.999$ increases the photon flux factor from 0.01 to roughly 0.5. Therefore, the $\gamma p$ cross sections given by ZEUS can be approximately converted to $ep$ cross sections by multiplying by 0.5.

\bibliography{ref} 

@article{MINERvA:2023avz,
    author = "Cai, T. and others",
    collaboration = "MINERvA",
    title = "{Measurement of the axial vector form factor from antineutrino\textendash{}proton scattering}",
    reportNumber = "FERMILAB-PUB-23-033-CSAID-ND-QIS",
    doi = "10.1038/s41586-022-05478-3",
    journal = "Nature",
    volume = "614",
    number = "7946",
    pages = "48--53",
    year = "2023"
}

@article{AguilarArevalo:2010MiniBooNECCQE,
  author  = {Aguilar-Arevalo, A. A. and others},
  collaboration = {MiniBooNE},
  title   = {First Measurement of the Muon Neutrino Charged Current Quasielastic Double Differential Cross Section},
  journal = {Phys. Rev. D},
  volume  = {81},
  pages   = {092005},
  year    = {2010},
  doi     = {10.1103/PhysRevD.81.092005},
  eprint  = {1002.2680},
  archivePrefix = {arXiv}
}

@article{Chen:2024ksq,
    author = "Chen, Yi",
    title = "{Nucleon relativistic weak-neutral axial-vector four-current distributions}",
    eprint = "2411.12521",
    archivePrefix = "arXiv",
    primaryClass = "hep-ph",
    doi = "10.1007/JHEP04(2025)132",
    journal = "JHEP",
    volume = "04",
    pages = "132",
    year = "2025"
}

@article{Graczyk:2021oyl,
    author = "Graczyk, Krzysztof M. and Kowal, Beata E.",
    title = "{Model dependence of the polarization asymmetries in weak pion production off the nucleon}",
    eprint = "2106.11383",
    archivePrefix = "arXiv",
    primaryClass = "hep-ph",
    doi = "10.1103/PhysRevD.104.033005",
    journal = "Phys. Rev. D",
    volume = "104",
    number = "3",
    pages = "033005",
    year = "2021"
}

@article{Rathmann:2025jgp,
    author = "Rathmann, F. and others",
    title = "{Eliminating beam-induced depolarizing effects in the hydrogen jet target for high-precision proton beam polarimetry at the Electron-Ion Collider}",
    eprint = "2508.01366",
    archivePrefix = "arXiv",
    primaryClass = "physics.acc-ph",
    month = "8",
    year = "2025"
}

@article{Huang:2025gqx,
    author = "Huang, H. and Hock, K. and Meot, F. and Ptitsyn, V. and Ranjbar, V. and Schoefer, V. and Morozov, V.",
    title = "{Hadron Polarization in the Electron Ion Collider}",
    doi = "10.22323/1.472.0022",
    journal = "PoS",
    volume = "PSTP2024",
    pages = "022",
    year = "2025"
}

@article{JeffersonLabHallA:1999epl,
    author = "Jones, M. K. and others",
    collaboration = "Jefferson Lab Hall A",
    title = "{$G_{Ep}/G_{Mp}$ ratio by
polarization transfer in $\vec ep\rightarrow e\vec p$}",
    eprint = "nucl-ex/9910005",
    archivePrefix = "arXiv",
    reportNumber = "JLAB-PHY-00-71",
    doi = "10.1103/PhysRevLett.84.1398",
    journal = "Phys. Rev. Lett.",
    volume = "84",
    pages = "1398--1402",
    year = "2000"
}

@article{Graczyk:2023lrm,
    author = "Graczyk, Krzysztof M. and Kowal, Beata E.",
    title = "{Neutral current neutrino and antineutrino scattering off the polarized nucleon}",
    eprint = "2307.00661",
    archivePrefix = "arXiv",
    primaryClass = "hep-ph",
    doi = "10.1103/PhysRevD.108.093002",
    journal = "Phys. Rev. D",
    volume = "108",
    number = "9",
    pages = "093002",
    year = "2023"
}

@article{Graczyk:2019xwg,
    author = "Graczyk, Krzysztof M. and Kowal, Beata E.",
    title = "{Spin asymmetries in quasielastic charged current neutrino-nucleon scattering}",
    eprint = "1912.00064",
    archivePrefix = "arXiv",
    primaryClass = "hep-ph",
    doi = "10.1103/PhysRevD.101.073002",
    journal = "Phys. Rev. D",
    volume = "101",
    number = "7",
    pages = "073002",
    year = "2020"
}

@article{T2K:2020CC0piOC,
  author  = {Abe, K. and others},
  collaboration = {T2K},
  title   = {Simultaneous measurement of the muon neutrino charged-current cross section on oxygen and carbon without pions in the final state at T2K},
  journal = {Phys. Rev. D},
  volume  = {101},
  number  = {11},
  pages   = {112004},
  year    = {2020},
  doi     = {10.1103/PhysRevD.101.112004},
  eprint  = {2004.05434},
  archivePrefix = {arXiv}
}

@article{Choi:1993PRL_AxialPionElectroprod,
  author  = {Choi, S. and Estenne, V. and Bardin, G. and De Botton, N. and Fournier, G. and Guichon, P. A. M. and Marchand, C. and Marroncle, J. and Martino, J. and Miller, J. and others},
  title   = {Axial and pseudoscalar nucleon form factors from low energy pion electroproduction},
  journal = {Phys. Rev. Lett.},
  volume  = {71},
  number  = {24},
  pages   = {3927--3930},
  year    = {1993},
  doi     = {10.1103/PhysRevLett.71.3927}
}

@article{Aschenauer:2025mku,
    author = "Aschenauer, Elke-Caroline and Bazilevsky, Alexander and Jentsch, Alexander and Kim, Jihee and Kiselev, Alexander and Page, Brian and Tu, Zhoudunming and Ullrich, Thomas and Wong, Cheuk-Ping",
    title = "{Tagging efficiency study of incoherent diffractive vector meson production at the second interaction region at the Electron-Ion Collider}",
    eprint = "2501.12410",
    archivePrefix = "arXiv",
    primaryClass = "physics.ins-det",
    doi = "10.1103/PhysRevD.111.072013",
    journal = "Phys. Rev. D",
    volume = "111",
    number = "7",
    pages = "072013",
    year = "2025"
}

@article{Sidoretti:2025tja,
    author = "Sidoretti, E. and others",
    title = "{The G-RWELL for ePIC Endcap Tracking}",
    doi = "10.1088/1748-0221/20/06/C06059",
    journal = "JINST",
    volume = "20",
    number = "06",
    pages = "C06059",
    year = "2025"
}

@article{Sidoretti:2025xeh,
    author = "Sidoretti, E. and others",
    title = "{The hybrid {\ensuremath{\mu}}-RWELL for ePIC Endcap tracking}",
    doi = "10.1016/j.nima.2025.170622",
    journal = "Nucl. Instrum. Meth. A",
    volume = "1080",
    pages = "170622",
    year = "2025"
}

@article{Philip:2024opz,
    author = "Philip, Olivier and others",
    title = "{Novel 4x4 SiPM array readout with integrated preamplification stage, optimized for the PWO detectors of the EIC EEEMCal}",
    doi = "10.1016/j.nima.2024.169278",
    journal = "Nucl. Instrum. Meth. A",
    volume = "1063",
    pages = "169278",
    year = "2024"
}

@article{Mkrtchyan:2025euu,
    author = "Mkrtchyan, Hamlet and Marukyan, Hrachya and Mkrtchyan, Arthur and Shahinyan, Albert and Tadevosyan, Vardan and Voskanyan, Hakob and Movsisyan, Aram and Hoghmrtsyan, Arthur",
    title = "{Lead Tungstate Electromagnetic Calorimeter Prototype Built in AANL for EIC}",
    doi = "10.1051/epjconf/202532000049",
    journal = "EPJ Web Conf.",
    volume = "320",
    pages = "00049",
    year = "2025"
}

@article{Zhang:2025ijg,
    author = "Zhang, Weibin and Preins, Sean and Huang, Jiajun and Paul, Sebouh J. and Milton, Ryan and Rodriguez, Miguel and Carney, Peter and Tsiao, Ryan and Abdelkadous, Yousef and Arratia, Miguel",
    title = "{First-ever deployment of a SiPM-on-tile calorimeter in a collider: a parasitic test with 200 GeV p p collisions at RHIC}",
    eprint = "2501.08586",
    archivePrefix = "arXiv",
    primaryClass = "physics.ins-det",
    doi = "10.1088/1748-0221/20/06/P06029",
    journal = "JINST",
    volume = "20",
    number = "06",
    pages = "P06029",
    year = "2025"
}

@article{Gardner:2023lly,
    author = "Gardner, Simon and Glazier, Derek I. and Livingston, Kenneth and Maneuski, Dzmitry and Sokhan, Daria and Adam, Jaroslav",
    title = "{Pixel-based tracking detectors for a Low Q2 Tagger at EIC -- status report}",
    eprint = "2305.02079",
    archivePrefix = "arXiv",
    primaryClass = "physics.ins-det",
    month = "5",
    year = "2023"
}

@inproceedings{Liu:2015cpe,
    author = {Liu, Yong and B{\"u}scher, Volker and Caudron, Julien and Chau, Phi and Krause, Sascha and Masetti, Lucia and Sch{\"a}fer, Ulrich and Spreckels, Rouven and Tapprogge, Stefan and Wanke, Rainer},
    title = "{A Design of Scintillator Tiles Read Out by Surface-Mounted SiPMs for a Future Hadron Calorimeter}",
    booktitle = "{2014 IEEE Nuclear Science Symposium and Medical Imaging Conference and 21st Symposium on Room-Temperature Semiconductor X-ray and Gamma-ray Detectors}",
    eprint = "1512.05900",
    archivePrefix = "arXiv",
    primaryClass = "physics.ins-det",
    doi = "10.1109/NSSMIC.2014.7431118",
    month = "12",
    year = "2015"
}

@article{H1:2009cml,
    author = "Aaron, F. D. and others",
    collaboration = "H1",
    title = "{Diffractive Electroproduction of rho and phi Mesons at HERA}",
    eprint = "0910.5831",
    archivePrefix = "arXiv",
    primaryClass = "hep-ex",
    reportNumber = "DESY-09-093",
    doi = "10.1007/JHEP05(2010)032",
    journal = "JHEP",
    volume = "05",
    pages = "032",
    year = "2010"
}

@article{ZEUS:2008hcd,
    author = "Chekanov, S. and others",
    collaboration = "ZEUS",
    title = "{A Measurement of the Q**2, W and t dependences of deeply virtual Compton scattering at HERA}",
    eprint = "0812.2517",
    archivePrefix = "arXiv",
    primaryClass = "hep-ex",
    reportNumber = "DESY-08-132",
    doi = "10.1088/1126-6708/2009/05/108",
    journal = "JHEP",
    volume = "05",
    pages = "108",
    year = "2009"
}

@article{Kumericki:2009uq,
    author = "Kumeri{\v{c}}ki, Kresimir and Mueller, Dieter",
    title = "{Deeply virtual Compton scattering at small $x_B$ and the access to the GPD H}",
    eprint = "0904.0458",
    archivePrefix = "arXiv",
    primaryClass = "hep-ph",
    doi = "10.1016/j.nuclphysb.2010.07.015",
    journal = "Nucl. Phys. B",
    volume = "841",
    pages = "1--58",
    year = "2010"
}

@article{H1:2005gdw,
    author = "Aktas, A. and others",
    collaboration = "H1",
    title = "{Measurement of deeply virtual compton scattering at HERA}",
    eprint = "hep-ex/0505061",
    archivePrefix = "arXiv",
    reportNumber = "DESY-05-065",
    doi = "10.1140/epjc/s2005-02345-3",
    journal = "Eur. Phys. J. C",
    volume = "44",
    pages = "1--11",
    year = "2005"
}

@article{Magnan:2017exp,
    author = "Magnan, A. M.",
    editor = "Govoni, Pietro and Lehmann, Giovanna and Marrocchesi, Pier Simone and Navarria, Francesco-Luigi and Paganoni, Marco and Perrotta, Andrea and Rovelli, Tiziano",
    collaboration = "CMS",
    title = "{HGCAL: a High-Granularity Calorimeter for the endcaps of CMS at HL-LHC}",
    doi = "10.1088/1748-0221/12/01/C01042",
    journal = "JINST",
    volume = "12",
    number = "01",
    pages = "C01042",
    year = "2017"
}

@article{Simon:2010hf,
    author = "Simon, F. and Soldner, C.",
    title = "{Uniformity Studies of Scintillator Tiles directly coupled to SiPMs for Imaging Calorimetry}",
    eprint = "1001.4665",
    archivePrefix = "arXiv",
    primaryClass = "physics.ins-det",
    reportNumber = "MPP-2010-10",
    doi = "10.1016/j.nima.2010.03.142",
    journal = "Nucl. Instrum. Meth. A",
    volume = "620",
    pages = "196--201",
    year = "2010"
}

@article{Blazey:2009zz,
    author = "Blazey, G. and others",
    title = "{Directly Coupled Tiles as Elements of a Scintillator Calorimeter with MPPC Readout}",
    reportNumber = "FERMILAB-PUB-09-585-E",
    doi = "10.1016/j.nima.2009.03.253",
    journal = "Nucl. Instrum. Meth. A",
    volume = "605",
    pages = "277--281",
    year = "2009"
}

@article{Atoian:2025dib,
    author = "Atoian, Grigor and others",
    title = "{Realizing the Scientific Program with Polarized Ion Beams at EIC}",
    eprint = "2510.10794",
    archivePrefix = "arXiv",
    primaryClass = "nucl-ex",
    month = "10",
    year = "2025"
}

@article{Nunes:2022cmc,
    author = "Nunes, Ana Sofia and Aschenauer, Elke Caroline and Eyser, Oleg and Schmidke, William",
    title = "{Hadron Polarimetry for the Electron-Ion Collider}",
    doi = "10.21468/SciPostPhysProc.8.163",
    journal = "SciPost Phys. Proc.",
    volume = "8",
    pages = "163",
    year = "2022"
}

@article{Meyer:2025rzh,
    author = "Meyer, A. S. and others",
    title = "{The Nucleon Axial Form Factor from Elementary Target Data}",
    eprint = "2512.14097",
    archivePrefix = "arXiv",
    primaryClass = "hep-ex",
    reportNumber = "LLNL-JRNL-2014317, FERMILAB-PUB-25-0912-LBNF-T",
    month = "12",
    year = "2025"
}

@article{Sefkow:2018rhp,
    author = "Sefkow, Felix and Simon, Frank",
    collaboration = "CALICE",
    title = "{A highly granular SiPM-on-tile calorimeter prototype}",
    eprint = "1808.09281",
    archivePrefix = "arXiv",
    primaryClass = "physics.ins-det",
    doi = "10.1088/1742-6596/1162/1/012012",
    journal = "J. Phys. Conf. Ser.",
    volume = "1162",
    number = "1",
    pages = "012012",
    year = "2019"
}

@misc{CFNS2025FixedTargetEIC,
  title        = {Exploring a Fixed-Target Program at the EIC: Feasibility and Physics Opportunities},
  howpublished = {CFNS Workshop, Stony Brook University/Online},
  month        = sep,
  year         = {2025},
  note         = {Sept.\ 29--30, 2025},
  url          = {https://indico.cfnssbu.physics.sunysb.edu/event/496/}
}

@misc{EIC_MasterParameter_Table_2023,
  title        = {Accelerator Master Parameter Table: Colliding Beam Parameters},
  author       = {{Electron-Ion Collider Project}},
  howpublished = {\url{https://eic.jlab.org/Parameters/index.html}},
  note         = {Tables 1--2, rows ``RMS divergence'' (H,V). Last updated 23 Jan 2023. Accessed 29 Oct 2025},
  year         = {2023},
  month        = {Jan}
}

@article{AbdulKhalek:2021gbh,
    author = "Abdul Khalek, R. and others",
    title = "{Science Requirements and Detector Concepts for the Electron-Ion Collider}: {EIC Yellow Report}",
    eprint = "2103.05419",
    archivePrefix = "arXiv",
    primaryClass = "physics.ins-det",
    reportNumber = "BNL-220990-2021-FORE, JLAB-PHY-21-3198, LA-UR-21-20953",
    doi = "10.1016/j.nuclphysa.2022.122447",
    journal = "Nucl. Phys. A",
    volume = "1026",
    pages = "122447",
    year = "2022"
}

@misc{Eljen_EJ200_204_208_212,
  author = {Eljen Technology},
  title  = {EJ-200, EJ-204, EJ-208, EJ-212 Plastic Scintillators: Data Sheet},
  year   = {n.d.},
  url    = {https://eljentechnology.com/products/plastic-scintillators/ej-200-ej-204-ej-208-ej-212},
  note   = {Rise times 0.7–1.0 ns; decay times 1.8 ns (EJ-204) and 3.3 ns (EJ-208). Accessed 2025-10-27.}
}

@article{Houzvicka:2022pny,
    author = "Hou{\v{z}}vi{\v{c}}ka, J. and Pol{\'a}k, J. and S{\'y}korov{\'a}, S. and Horn, T. and Berdnikov, V. and Philip, O. and Shestakova, I.",
    title = "{PbWO4 (Lead tungstate) scintillator crystal production and evaluation for the electron ion collider electron endcap electromagnetic calorimeter (EEEMcal)}",
    doi = "10.1117/12.2632788",
    journal = "Proc. SPIE Int. Soc. Opt. Eng.",
    volume = "12241",
    pages = "1224104",
    year = "2022"
}

@article{Chatterjee:2024zrn,
    author = "Chatterjee, Chandradoy",
    collaboration = "ePIC",
    title = "{Particle Identification with the ePIC detector at the EIC}",
    eprint = "2410.20410",
    archivePrefix = "arXiv",
    primaryClass = "physics.ins-det",
    doi = "10.22323/1.469.0266",
    journal = "PoS",
    volume = "DIS2024",
    pages = "266",
    year = "2025"
}

@article{Li:2023hjr,
    author = "Li, Xuan",
    title = "{Latest Vertex and Tracking Detector Developments for the Future Electron{\textendash}Ion Collider}",
    eprint = "2305.15593",
    archivePrefix = "arXiv",
    primaryClass = "physics.ins-det",
    reportNumber = "LA-UR-22-32366",
    doi = "10.7566/JPSCP.42.011032",
    journal = "JPS Conf. Proc.",
    volume = "42",
    pages = "011032",
    year = "2024"
}

@article{Gonella:2024cxk,
    author = "Gonella, Laura",
    collaboration = "ePIC Silicon Vertex Tracker Detector Subsystem",
    title = "{Development of a Silicon Vertex and Tracking Detector for the Electron-Ion Collider}",
    doi = "10.22323/1.448.0038",
    journal = "PoS",
    volume = "VERTEX2023",
    pages = "038",
    year = "2024"
}

@article{Bhattacharya:2023ddr,
    author = "Bhattacharya, D. S. and Cisbani, E. and Chatterjee, C. and Dalla Torre, S. and Dilks, C. and Kiselev, A. and Klest, H. and Preghenella, R. and Vossen, A.",
    title = "{Simulation studies related to the particle identification by the forward and backward RICH detectors at Electron Ion Collider}",
    eprint = "2301.08334",
    archivePrefix = "arXiv",
    primaryClass = "physics.ins-det",
    doi = "10.1016/j.nima.2023.168591",
    journal = "Nucl. Instrum. Meth. A",
    volume = "1056",
    pages = "168591",
    year = "2023"
}

@article{Liesenfeld:1999PLB_AxialFromPeepiN,
  author  = {Liesenfeld, A. and others},
  title   = {A measurement of the axial form factor of the nucleon by the $p(e,e'\pi^+)n$ reaction at low momentum transfer},
  journal = {Phys. Lett. B},
  volume  = {468},
  pages   = {20--26},
  year    = {1999},
  doi     = {10.1016/S0370-2693(99)01204-4}
}

@article{Formaggio:2012cpf,
    author = "Formaggio, J. A. and Zeller, G. P.",
    title = "{From eV to EeV: Neutrino Cross Sections Across Energy Scales}",
    eprint = "1305.7513",
    archivePrefix = "arXiv",
    primaryClass = "hep-ex",
    reportNumber = "FERMILAB-PUB-12-785-E",
    doi = "10.1103/RevModPhys.84.1307",
    journal = "Rev. Mod. Phys.",
    volume = "84",
    pages = "1307--1341",
    year = "2012"
}

@article{ZEUS:2002gig,
    author = "Chekanov, S. and others",
    collaboration = "ZEUS",
    title = "{Leading neutron production in e+ p collisions at HERA}",
    eprint = "hep-ex/0205076",
    archivePrefix = "arXiv",
    reportNumber = "DESY-02-039",
    doi = "10.1016/S0550-3213(02)00439-X",
    journal = "Nucl. Phys. B",
    volume = "637",
    pages = "3--56",
    year = "2002"
}

@article{Ambrosino:2015ruf,
    author = "Ambrosino, F. and others",
    title = "{CHANTI: a Fast and Efficient Charged Particle Veto Detector for the NA62 Experiment at CERN}",
    eprint = "1512.00244",
    archivePrefix = "arXiv",
    primaryClass = "physics.ins-det",
    doi = "10.1088/1748-0221/11/03/P03029",
    journal = "JINST",
    volume = "11",
    number = "03",
    pages = "P03029",
    year = "2016"
}

@techreport{JLab-PR12-25-009,
  author       = {Averett, T. and Napolitano, J. and Wojtsekhowski, B. and Xiong, W., et. al},
  title        = {The Nucleon Axial-Vector Form Factor from the H($\vec{e}$, n)$\nu_e$ Reaction},
  institution  = {Thomas Jefferson National Accelerator Facility},
  address      = {Newport News, VA},
  number       = {PR12-25-009},
  type         = {Proposal},
  year         = {2025},
  month        = {July},
  note         = {Proposal to JLab PAC 53},
  url          = {https://indico.jlab.org/event/923/contributions/17312/}
}

@article{Pire:2021dad,
    author = "Pire, B. and Szymanowski, L. and Wagner, J.",
    title = "{Charged current electroproduction of a charmed meson at an electron-ion collider}",
    eprint = "2104.04944",
    archivePrefix = "arXiv",
    primaryClass = "hep-ph",
    reportNumber = "CPHT-RR028.042021",
    doi = "10.1103/PhysRevD.104.094002",
    journal = "Phys. Rev. D",
    volume = "104",
    number = "9",
    pages = "094002",
    year = "2021"
}

@article{Siddikov:2019ahb,
    author = "Siddikov, Marat and Schmidt, Ivan",
    title = "{Generalized Parton Distributions from charged current meson production}",
    eprint = "1904.04252",
    archivePrefix = "arXiv",
    primaryClass = "hep-ph",
    reportNumber = "USM-TH-360",
    doi = "10.1103/PhysRevD.99.116005",
    journal = "Phys. Rev. D",
    volume = "99",
    number = "11",
    pages = "116005",
    year = "2019"
}

@article{Kharzheev:2019tfk,
    author = "Kharzheev, Yu. N.",
    title = "{Radiation Hardness of Scintillation Detectors Based on Organic Plastic Scintillators and Optical Fibers}",
    doi = "10.1134/S1063779619010027",
    journal = "Phys. Part. Nucl.",
    volume = "50",
    number = "1",
    pages = "42--76",
    year = "2019"
}

@article{Tomalak:2020zlv,
    author = "Tomalak, Oleksandr",
    title = "{Axial and pseudoscalar form factors from charged current quasielastic neutrino-nucleon scattering}",
    eprint = "2008.03527",
    archivePrefix = "arXiv",
    primaryClass = "hep-ph",
    reportNumber = "FERMILAB-PUB-20-418-T",
    doi = "10.1103/PhysRevD.103.013006",
    journal = "Phys. Rev. D",
    volume = "103",
    number = "1",
    pages = "013006",
    year = "2021"
}

@article{ZEUS:2007knd,
    author = "Chekanov, S. and others",
    collaboration = "ZEUS",
    title = "{Leading neutron energy and pT distributions in deep inelastic scattering and photoproduction at HERA}",
    eprint = "hep-ex/0702028",
    archivePrefix = "arXiv",
    reportNumber = "DESY-07-011",
    doi = "10.1016/j.nuclphysb.2007.03.045",
    journal = "Nucl. Phys. B",
    volume = "776",
    pages = "1--37",
    year = "2007"
}

@article{Sjostrand:2006za,
    author = "Sjostrand, Torbjorn and Mrenna, Stephen and Skands, Peter Z.",
    title = "{PYTHIA 6.4 Physics and Manual}",
    eprint = "hep-ph/0603175",
    archivePrefix = "arXiv",
    reportNumber = "FERMILAB-PUB-06-052-CD-T, LU-TP-06-13",
    doi = "10.1088/1126-6708/2006/05/026",
    journal = "JHEP",
    volume = "05",
    pages = "026",
    year = "2006"
}

@article{LlewellynSmith1972,
  author  = {Llewellyn Smith, C. H.},
  title   = {Neutrino Reactions at Accelerator Energies},
  journal = {Phys. Rep.},
  volume  = {3},
  pages   = {261--379},
  year    = {1972},
  doi     = {10.1016/0370-1573(72)90010-5}
}

@article{PolyakovSchweitzer2018,
  author  = {Polyakov, Maxim V. and Schweitzer, Peter},
  title   = {Forces Inside Hadrons: Pressure, Surface Tension, Mechanical Radius, and All That},
  journal = {Int. J. Mod. Phys. A},
  volume  = {33},
  pages   = {1830025},
  year    = {2018},
  doi     = {10.1142/S0217751X18300259}
}

@article{Hofstadter1956,
  author  = {Hofstadter, Robert},
  title   = {Electron Scattering and Nuclear Structure},
  journal = {Rev. Mod. Phys.},
  volume  = {28},
  pages   = {214--254},
  year    = {1956},
  doi     = {10.1103/RevModPhys.28.214}
}

@article{Chen:2024oxx,
    author = "Chen, Yi and Li, Yang and Lorc\'e, C\'edric and Wang, Qun",
    title = "{Nucleon axial radius}",
    eprint = "2405.12943",
    archivePrefix = "arXiv",
    primaryClass = "hep-ph",
    doi = "10.1103/PhysRevD.110.L091503",
    journal = "Phys. Rev. D",
    volume = "110",
    number = "9",
    pages = "L091503",
    year = "2024"
}

@article{Goharipour:2025yxm,
    author = "Goharipour, Muhammad and Irani, Fatemeh and Amiri, M. H. and Fatehi, H. and Falahi, Behnam and Moradi, A. and Azizi, K.",
    collaboration = "MMGPDs",
    title = "{Can we determine the exact size of the nucleon?: A comprehensive study of different radii}",
    eprint = "2503.08847",
    archivePrefix = "arXiv",
    primaryClass = "hep-ph",
    doi = "10.1016/j.nuclphysb.2025.116962",
    journal = "Nucl. Phys. B",
    volume = "1017",
    pages = "116962",
    year = "2025"
}

@article{Alexandrou:2017hac,
    author = "Alexandrou, Constantia and Constantinou, Martha and Hadjiyiannakou, Kyriakos and Jansen, Karl and Kallidonis, Christos and Koutsou, Giannis and Vaquero Aviles-Casco, Alejandro",
    title = "{Nucleon axial form factors using $N_f$ = 2 twisted mass fermions with a physical value of the pion mass}",
    eprint = "1705.03399",
    archivePrefix = "arXiv",
    primaryClass = "hep-lat",
    reportNumber = "DESY-17-064",
    doi = "10.1103/PhysRevD.96.054507",
    journal = "Phys. Rev. D",
    volume = "96",
    number = "5",
    pages = "054507",
    year = "2017"
}

@article{RQCD:2019jai,
    author = {Bali, Gunnar S. and Barca, Lorenzo and Collins, Sara and Gruber, Michael and L{\"o}ffler, Marius and Sch{\"a}fer, Andreas and S{\"o}ldner, Wolfgang and Wein, Philipp and Weish{\"a}upl, Simon and Wurm, Thomas},
    collaboration = "RQCD",
    title = "{Nucleon axial structure from lattice QCD}",
    eprint = "1911.13150",
    archivePrefix = "arXiv",
    primaryClass = "hep-lat",
    doi = "10.1007/JHEP05(2020)126",
    journal = "JHEP",
    volume = "05",
    pages = "126",
    year = "2020"
}

@article{Haberzettl:2000sm,
    author = "Haberzettl, Helmut",
    title = "{Pion photoproduction and electroproduction and the partially conserved axial current}",
    eprint = "nucl-th/0005016",
    archivePrefix = "arXiv",
    doi = "10.1103/PhysRevLett.85.3576",
    journal = "Phys. Rev. Lett.",
    volume = "85",
    pages = "3576--3579",
    year = "2000"
}

@article{NuSTEC:2017hzk,
    author = "Alvarez-Ruso, L. and others",
    collaboration = "NuSTEC",
    title = "{NuSTEC White Paper: Status and challenges of neutrino{\textendash}nucleus scattering}",
    eprint = "1706.03621",
    archivePrefix = "arXiv",
    primaryClass = "hep-ph",
    reportNumber = "FERMILAB-PUB-17-195-ND-T, INT-PUB-17-020",
    doi = "10.1016/j.ppnp.2018.01.006",
    journal = "Prog. Part. Nucl. Phys.",
    volume = "100",
    pages = "1--68",
    year = "2018"
}

@article{Capitani:2017qpc,
    author = {Capitani, Stefano and Della Morte, Michele and Djukanovic, Dalibor and von Hippel, Georg M. and Hua, Jiayu and J{\"a}ger, Benjamin and Junnarkar, Parikshit M. and Meyer, Harvey B. and Rae, Thomas D. and Wittig, Hartmut},
    title = "{Isovector axial form factors of the nucleon in two-flavor lattice QCD}",
    eprint = "1705.06186",
    archivePrefix = "arXiv",
    primaryClass = "hep-lat",
    reportNumber = "CP3-Origins-2017-018, HIM-2017-03, MITP/17-029, TIFR/TH/17-21, CP3-ORIGINS-2017-018, MITP-17-029, TIFR-TH-17-21",
    doi = "10.1142/S0217751X1950009X",
    journal = "Int. J. Mod. Phys. A",
    volume = "34",
    number = "02",
    pages = "1950009",
    year = "2019"
}

@article{Alexandrou:2020okk,
    author = "Alexandrou, C. and others",
    title = "{Nucleon axial and pseudoscalar form factors from lattice QCD at the physical point}",
    eprint = "2011.13342",
    archivePrefix = "arXiv",
    primaryClass = "hep-lat",
    doi = "10.1103/PhysRevD.103.034509",
    journal = "Phys. Rev. D",
    volume = "103",
    number = "3",
    pages = "034509",
    year = "2021"
}

@article{Gupta:2017dwj,
    author = "Gupta, Rajan and Jang, Yong-Chull and Lin, Huey-Wen and Yoon, Boram and Bhattacharya, Tanmoy",
    title = "{Axial Vector Form Factors of the Nucleon from Lattice QCD}",
    eprint = "1705.06834",
    archivePrefix = "arXiv",
    primaryClass = "hep-lat",
    reportNumber = "LA-UR-17-23678",
    doi = "10.1103/PhysRevD.96.114503",
    journal = "Phys. Rev. D",
    volume = "96",
    number = "11",
    pages = "114503",
    year = "2017"
}

@article{Tsuji:2025quu,
    author = "Tsuji, Ryutaro and Aoki, Yasumichi and Ishikawa, Ken-Ichi and Kuramashi, Yoshinobu and Sasaki, Shoichi and Sato, Kohei and Shintani, Eigo and Watanabe, Hiromasa and Yamazaki, Takeshi",
    collaboration = "PACS",
    title = "{Investigating the axial structure of the nucleon based on large-volume lattice QCD at the physical point}",
    eprint = "2505.10998",
    archivePrefix = "arXiv",
    primaryClass = "hep-lat",
    reportNumber = "UTHEP-803, UTCCS-P-166, HUPD-2504, KEK-TH-2716",
    month = "5",
    year = "2025"
}

@article{Miller:2019DefiningRp,
  author  = {Miller, Gerald A.},
  title   = {Defining the proton radius: A unified treatment},
  journal = {Phys. Rev. C},
  volume  = {99},
  pages   = {035202},
  year    = {2019},
  doi     = {10.1103/PhysRevC.99.035202},
  eprint  = {1812.02714},
  archivePrefix = {arXiv}
}

@article{KaiserWeise:2024Sizes,
  author  = {Kaiser, Norbert and Weise, Wolfram},
  title   = {Sizes of the nucleon},
  journal = {Phys. Rev. C},
  volume  = {110},
  pages   = {015202},
  year    = {2024},
  doi     = {10.1103/PhysRevC.110.015202}
}

@article{Milton:2024bqv,
    author = "Milton, Ryan and Paul, Sebouh J. and Schmookler, Barak and Arratia, Miguel and Karande, Piyush and Angerami, Aaron and Acosta, Fernando Torales and Nachman, Benjamin",
    title = "{Design and simulation of a SiPM-on-tile ZDC for the future EIC, and its performance with graph neural networks}",
    eprint = "2406.12877",
    archivePrefix = "arXiv",
    primaryClass = "physics.ins-det",
    doi = "10.1016/j.nima.2025.170613",
    journal = "Nucl. Instrum. Meth. A",
    volume = "1079",
    pages = "170613",
    year = "2025"
}
\end{document}